\documentclass[a4paper]{article}
\usepackage{INTERSPEECH2019}
\usepackage{amsmath,graphicx,cite}
\usepackage{tipa}
\usepackage{color}

\def\mathbi#1{\textbf{\em #1}}

\makeatletter
\newcommand{\thickhline}{%
    \noalign {\ifnum 0=`}\fi \hrule height 1pt
    \futurelet \reserved@a \@xhline
}
\makeatother

\title{LSTM Acoustic Models Learn to Align and Pronounce with Graphemes}

\name{Arindrima Datta{\normalfont\textsuperscript{1}}, Guanlong Zhao{\normalfont\textsuperscript{*2}}\thanks{*Work done at Google NYC.}, Bhuvana Ramabhadran{\normalfont\textsuperscript{1}}, Eugene Weinstein{\normalfont\textsuperscript{1}}}
\address{
    \textsuperscript{1}Google, Inc., New York, NY, U.S.A.\\
    \textsuperscript{2}Texas A\&M University, College Station, TX, U.S.A.}
\email{\{arindrimadatta, bhuv, weinstein\}@google.com, gzhao@tamu.edu}

\begin{document}
\ninept
\maketitle
\begin{abstract}
Automated speech recognition coverage of the world's languages
continues to expand. However, standard phoneme based systems require handcrafted
lexicons that are difficult and expensive to obtain. To address this problem, we
propose a training methodology for a grapheme-based speech recognizer
that can be trained in a purely data-driven fashion. Built with LSTM
networks and trained with the cross-entropy loss, the grapheme-output
acoustic models we study are also extremely practical for real-world
applications as they can be
decoded with conventional ASR stack components such as language models
and FST decoders, and produce good quality audio-to-grapheme alignments
that are useful in many speech applications. We show that the
grapheme models are competitive in WER with their phoneme-output
counterparts when trained on large datasets, with the advantage that
grapheme models do not require explicit linguistic knowledge as an
input. We further compare the alignments generated by the phoneme and
grapheme models to demonstrate the quality of the pronunciations
learnt by them using four Indian languages that vary linguistically
in spoken and
written forms.

\end{abstract}
\section{Introduction}
\label{sec:intro}

Automated speech recognition (ASR) performance has seen rapid improvements in recent years.
Conventional speech recognizers are comprised of four main components:
acoustic, language, and pronunciation (lexicon) models, and search
decoders. The acoustic and language models are statistical models that
have been improved over the last few decades with newer algorithms and
increased training data. In contrast, the lexicon is usually manually generated using a
dictionary of human transcribed word pronunciations. Human-curated dictionaries suffer
from various challenges, such as the prohibitive cost involved
with acquiring pronunciations for the different dialects and accents of a language,
as well as for new languages.

To address these challenges, there have been several statistical approaches aimed at automated unit and lexicon
discovery from speech audio~\cite{ramabhadran1998acoustics,lu2013acoustic},
grapheme-to-phoneme (g2p) conversion~\cite{chen2003conditional} and
more recently with the use of Long Short-Term Memory (LSTM) networks~\cite{rao2015grapheme}. However, these models
still need to be trained on manually curated pronunciation dictionaries.

Grapheme models have also been used to address the challenges of acquiring lexica.
For instance, in the iARPA-sponsored BABEL program that focused
on the rapid development of ASR and keyword-spotting systems for low resource
languages, graphemic systems were widely
used~\cite{le2014developing, wang2015joint, golik2015multilingual,
  trmal2017kaldi, gales2015unicode}, and it was shown that
graphemic ASR systems can yield similar performance to the phonemic
systems. In this paper, we aim to build graphemic speech recognizers
with \emph{large scale speech data for high-resource languages}
but without the use of lexica.

LSTM cells can model varying contexts (``memory") and have been successful
in a number of sequence prediction tasks~\cite{chan2016listen,bahdanau2014neural,
chorowski2015attention}.
In ASR, grapheme LSTMs have been used with the Connectionist Temporal Classification
(CTC) loss~\cite{graves2006connectionist,graves2014towards}
in~\cite{rao2017multi, cui2017knowledge, rosenberg2017end}, and with lattice-free MMI
objective in~\cite{wang2018phonetic} to obtain competitive performance in speech recognition.
Encoder-decoder based grapheme
models~\cite{chan2016listen,chorowski2015attention} doing end-to-end (E2E)
speech recognition have also been proven to be
competitive with conventional models.

As these models become increasingly popular, an understanding of the
alignment they produce between speech frames and linguistic elements such as
phonemes, graphemes, or words is important for not only ASR but also tasks
such as keyword spotting~\cite{sun2016max, audhkhasi2017end},
grapheme-to-phoneme conversion~\cite{jyothi2017low},
captioning~\cite{soltau2016neural,federico2014automatic}
and speech synthesis~\cite{wang2017tacotron}.
While graphemic models have been studied in the traditional HMM-GMM
framework~\cite{killer2003grapheme,sung2009revisiting}, the literature
does not offer us a thorough understanding of the quality of graphemic
alignments produced by these models.
Recently~\cite{zhang2018exploration} explored the use of word alignments produced by
direct acoustics-to-word models trained with a cross-entropy objective.
However, word-level modeling of speech is problematic due to out-of-vocabulary
(OOV) issues ~\cite{soltau2016neural,audhkhasi2017direct},
which we can avoid by using graphemes instead.

Motivated by the success of graphemic LSTM neural networks, in this paper we study
their ability to implicitly learn alignments and pronunciations
while being trained to recognize speech.
\emph{Using four Indian languages consiting of 2K-18K hours of
speech data, and with varying phonetic orthography and accents,
we demonstrate the ability of LSTMs to align grapheme sequences with
spoken audio}. Our cross-entropy (CE) training of these models
allows alignment quality to be explicitly optimized as part of the
training objective, resulting in high-quality acoustic-to-grapheme
alignments.
Furthermore, we analyze and compare the quality of alignments between the
grapheme and phoneme models, and attempt to \emph{understand the pronunciation-modeling
capability of these graphemic LSTMs}. To the best of our knowledge, this is one of the first pieces of work on large
scale data that specifically attempts to interpret the pronunciations and alignments
learnt by graphemic neural networks.

Our proposed training methodology also \emph{allows flat starting, and is purely data-driven}.
Therefore if offers the benefit of creating speech recognizers for new languages,
an use-case where grapheme models have shown the biggest potential.
Our grapheme acoustic models (AMs) trained with large-scale speech data,
achieve similar WERs in speech recognition to that of
phoneme models while avoiding the use of manually-curated lexicons.
Additionally, due to their \emph{ability to be
combined with conventional speech-modeling elements such as language models
and finite-state transducer (FST) decoders, they offer a
straightforward launch path}, and are thus advantageous over grapheme-output E2E models.

\section{Methods}
\label{sec:methods}

In this section, we describe the graphemic lexicon and the GMM-based graphemic alignments used
to flat start the graphemic LSTM acoustic models.

\subsection{Graphemic Lexicon}
\label{ssec:lexicon}
A graphemic lexicon is a mapping between graphemes and words with a special
``\textless space\textgreater" grapheme acting as the word boundary.
The set of graphemes for a language are generated
by enumerating the characters empirically observed in the training data,
after removing the very rare graphemes (\textless 10 occurrences) and those without any
clear acoustic realization (such as emojis). Utterances containing the excluded
graphemes are dropped from the training data.
The number of graphemes used in each language is
presented in the first column of Table \ref{table:graphemes}.

\subsection{LSTM and Cross-Entropy Loss}
\label{ssec:ce}

LSTMs can model long-range temporal dependencies, which make them
particularly suited for ASR. In real-time streaming ASR applications such as
voice search, we prefer uni-directional over bi-directional LSTM
\cite{rao2017multi} because the former is able to perform recognition in an online fashion.
The LSTM AM takes the current feature sequence $\mathbi{x}=(x_1, \ldots, x_T)$ as inputs and
estimates the output posteriors over a pre-defined label set, $\mathbi{l}=(l_1, \ldots, l_T)$. The LSTM
parameters can thus be estimated by maximizing the
cross-entropy (CE) loss \cite{sak2015learning} on all frames of input utterance $\mathbi{x}$ with its
corresponding frame-level alignment $\mathbi{l}$.

\begin{align}
    L_{CE} = -\sum_{(\mathbi{x},~\mathbi{l})}\sum_{l,~t}\delta(l, l_t)y_l^{t},
\end{align}
\noindent
where $\delta(\cdot,\cdot)$ is the Kronecker delta,
and $y_l^{t}$ is the network output activation for label $l$ at time $t$.

Cross-entropy trained LSTM acoustic models (AMs) are particularly appealing due to the simplicity and power
of this loss function~\cite{rubinstein2013cross},
a learning algorithm that lends itself well to efficient
implementation for diverse model architectures and
hardware~\cite{graves2013hybrid, wiesler2013investigations, vesely2010parallel}

\subsection{GMM Alignments}
\label{ssec:gmm_align}

Conventional phoneme-based systems use an existing speech recognizer to
generate forced-alignments to get the boundaries of the phone segments, thus,
assigning phoneme labels to the frames. Similarly, for CE-based grapheme AMs,
we need frame-level graphemic alignments to provide the training labels.
This approach of flat starting used in conventional phoneme-based models (e.g.,\cite{senior2014gmm})
can be applied to grpahemes as well. Flat starting is particularly useful for
training speech recognizers for new languages, which is
often the use-case for a grapheme-based model.

Since GMMs are easy to train and have been studied extensively in the literature,
we use them to generate initial alignments for the grapheme systems, using the
training methodology similar to \cite{mo2017}.
The speech signal is segmented evenly according to the target
graphemic sequences, and the GMMs are trained to
predict 3-state HMM context-independent graphemes using the
Expectation Maximization (EM) algorithm.
The input features are perceptual linear predictive features (PLPs) with
deltas and delta-deltas \cite{mo2017}, and the GMMs are trained with
14 mixtures per grapheme
on less than 6\% of the data for each language.

\begin{table}[!t]
\caption{Number of graphemes and phonemes per language}
\label{table:graphemes}
\centering
\begin{tabular}{|c||c|c|}
\thickhline
Language     &\# graphemes  &\# phonemes\\ \hline
Bengali & 96 & 39 \\
Tamil   & 90 & 35 \\
Hindi   & 114 & 52 \\
English & 56 & 53 \\ \thickhline
\end{tabular}
\end{table}

\begin{table}[!t]
\caption{Training and test data size}
\label{table:data}
\centering
\begin{tabular}{|c||c|c|}
\thickhline
Language     &\# hour (train) &\# hours (test)\\\hline
Bengali      & 4.5K   &4.1  \\
Tamil        & 1.9K   &6.4  \\
Hindi        & 18K    &7.2  \\
English      & 18K    &2.4  \\ \thickhline
\end{tabular}
\end{table}

\begin{table}[!t]
\caption{WER : \textbf{P}honeme and \textbf{G}rapheme models after \textbf{CE} and \textbf{sMBR} training}
\label{table:wer_ce_smbr}
\centering
\begin{tabular}{|c||cc|cc|}
\thickhline
Language & CE-P & CE-G & sMBR-P & sMBR-G \\ \hline
Bengali  & 36.7 & 41.7  & 29.2  & 31.0(+1.8)       \\
Tamil    & 41.5 & 43.2  & 30.3  & 31.9(+1.7)        \\
Hindi    & 33.1 & 38.1  & 25.6  & 28.1(+2.5)        \\
English  & 23.8 & 24.8  & 15.2  & 18.6(+3.4)       \\ \thickhline
\end{tabular}
\end{table}

\begin{table}[!t]
\caption{WER: Acoustic models trained with \textbf{GMM} vs \textbf{CE-LSTM} alignments with \textbf{P}honeme and \textbf{G}rapheme targets}
\label{table:wer_ce_realignment}
\centering
\begin{tabular}{|c||cc|cc|}
\thickhline
Language     & GMM-P & GMM-G & CE-P & CE-G \\ \hline
Bengali & 36.7 & 41.7(+5.0)  & 32.9  & 35.7(+2.8)\\
Tamil  & 41.5 & 43.2(+1.7)  & 36.8  & 37.1(+0.3)\\
Hindi  & 33.1 & 38.1(+5.0)  & 30.7  & 34.2(+3.5)\\
English & 23.8 & 24.8(+1.0)  & 22.6  & 23.1(+0.5)\\ \thickhline
\end{tabular}
\end{table}

\section{Experimental Setup and Results}
\label{sec:experiments}

For each of the four languages, Bengali, Tamil, Hindi and Indian English,
we used the same amount of data for training the grapheme and phoneme models.
Our training data consisted of anonymized,
human-transcribed utterances representative of Google's traffic.
The amount of transcribed data per language is tabulated in Table \ref{table:data}.
To achieve noise robustness the training data
is augmented with varying degrees of noise and reverberation such that the overall SNR is
between 0dB and 30dB, and the average SNR is 12dB \cite{Chanwoo17}.
The noise sources are from YouTube and daily life noisy
environmental recordings.

The phoneme models were trained with the same recipe (Section \ref{sec:methods}) but using phoneme targets instead.
All experiments used 80-dimensional log-mel features,
computed with a 25ms window and shifted every 10ms. These features were stacked with 7 frames to the left and
down-sampled to 30ms frame rate, following \cite{sak2015fast}.

We used 5-layer$\times$768 uni-directional LSTMs for cross-entropy training,
which was followed by state-level minimum Bayes risk (sMBR) training \cite{kingsbury2009lattice}, using the same model architecture.
Standard FST-based beam-search decoders with 5-gram language models from the target languages were used for decoding.

Since Indian languages frequently contain Latin characters in the transcripts, all our models were
evaluated after normalization for transliteration errors (\textit{transliterated WER} \cite{emond2018}).
Or in other words, if the model correctly decoded a word in Latin
alphabet but the reference was in the native script (or vice versa), this
was not considered an error.

Table \ref{table:wer_ce_smbr} presents the WER of Phoneme and Grapheme models after
CE and sMBR training, and we see an average improvement in performance of
$\sim$27.5\% for all models after sequence training with sMBR criteria.
Using the alignments generated by the respective CE models, we re-aligned and
re-trained both phoneme and grapheme models.
Both, phoneme and grapheme models showed a relative improvement between 5 to 15\%;
however, the gap between them also narrowed (Table \ref{table:wer_ce_realignment}).
The suffixes \textbf{P} and \textbf{G} refer to \textbf{P}honeme and \textbf{G}rapheme models in both the tables.

\begin{figure}[!t]
    \begin{center}
        \includegraphics[trim=90 200 80 200, clip, width=3in]{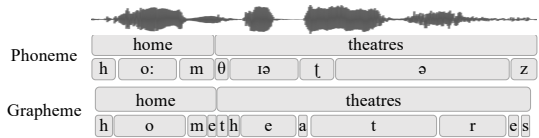}
    \caption{Grapheme alignment example. From top to bottom: speech waveform, word alignment by the phoneme model, phoneme alignment, word alignment by the grapheme model, and grapheme alignment.}
    \label{fig:align_example}
    \end{center}
\end{figure}

\begin{figure}[!t]
    \begin{center}
        \includegraphics[trim=90 35 45 20, clip, width=3in]{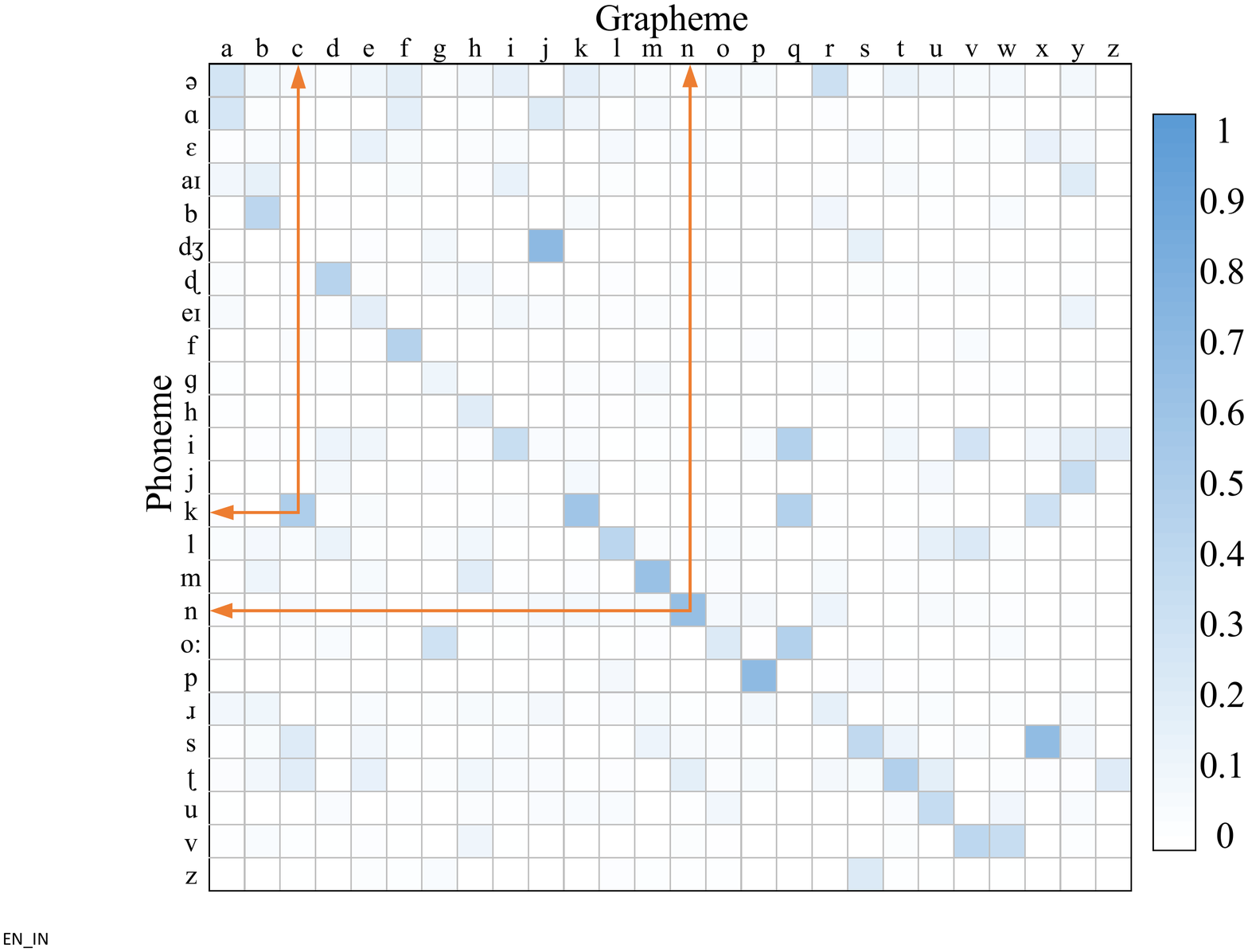}
    \caption{Alignment confusion matrix for Indian English. Only a subset of phonemes/graphemes are kept for better visualization.}
    \label{fig:cmat_en_in}
    \end{center}
\end{figure}

\begin{figure}[!t]
    \begin{center}
        \includegraphics[trim=90 35 45 20, clip, width=3in]{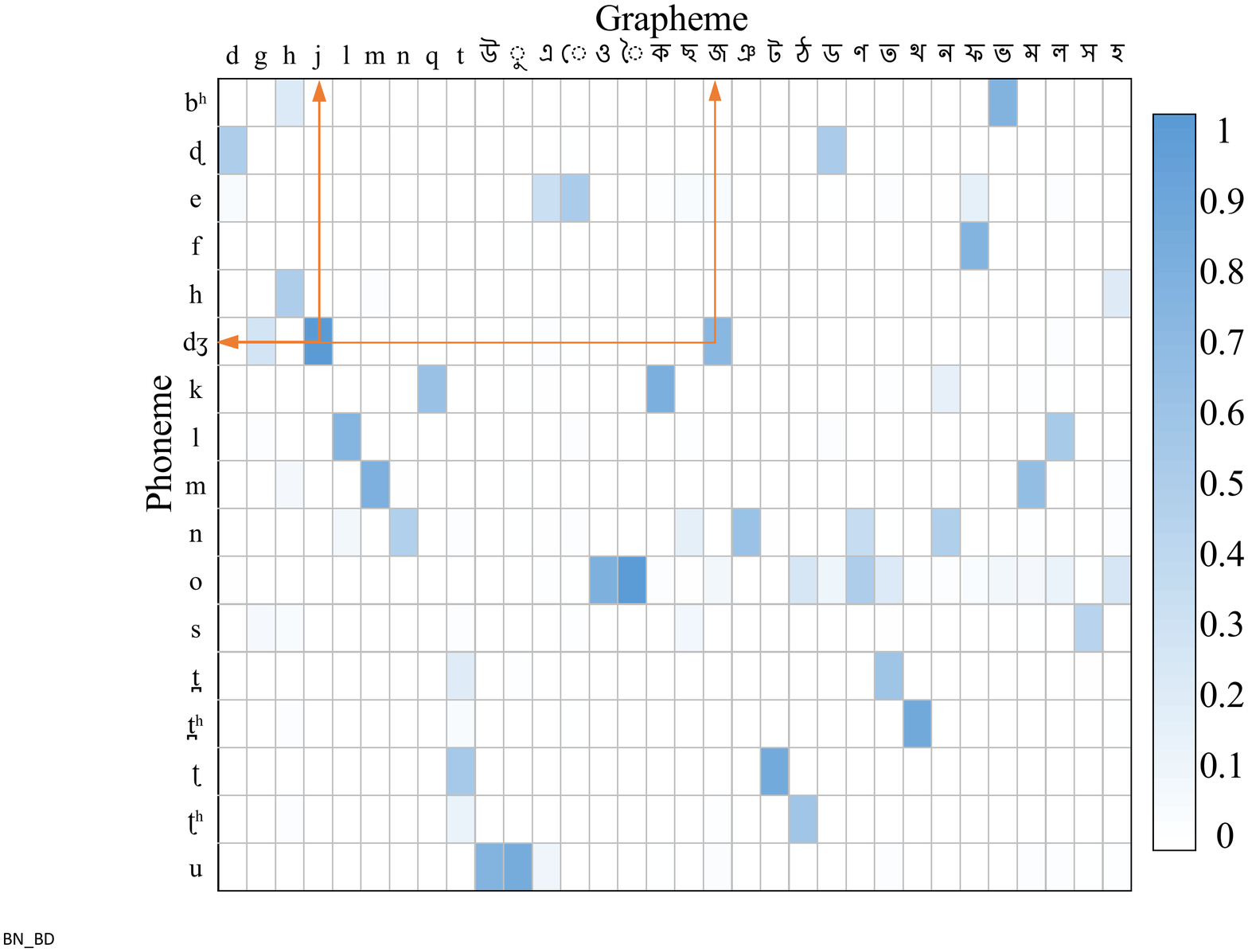}
    \caption{Alignment confusion matrix for Bengali. Only a subset of phonemes/graphemes are kept for better visualization.}
    \label{fig:cmat_bn_bd}
    \end{center}
\end{figure}

\begin{figure}[!b]
    \begin{center}
        \includegraphics[trim=100 195 55 180, clip, width=3in]{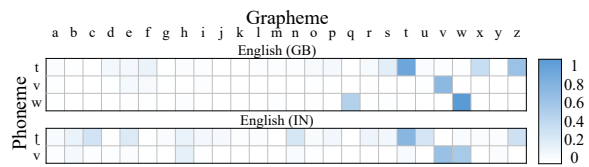}
    \caption{Some examples of differences in the alignment confusion matrices between British (GB) and Indian (IN) English. Note that in our phoneme set, English (IN) does not have the phoneme ``w."}
    \label{fig:cmat_en_gb_in}
    \end{center}
\end{figure}

\section{Alignments: Phoneme vs Grapheme}
\label{ssec:comparison}

In order to study the alignments generated by the graphemic models,
we compared these with phonemic alignments. An example of comparison is presented
in Fig.~\ref{fig:align_example}. More generally, we generated the graphemic and
phonemic alignments on the same set of utterances, which is presented in confusion matrices
for English in Fig.~\ref{fig:cmat_en_in}, and for Bengali in Fig.~\ref{fig:cmat_bn_bd}.
The matrices are normalized by grapheme distribution (i.e., by columns). We can see that
for Bengali, a language with strong phoneme-grapheme correspondence, the graphemes
match their corresponding phonemes.
For example, Bengali symbol ``\includegraphics[trim=200 120 260 120, height=\fontcharht\font`\B,clip]{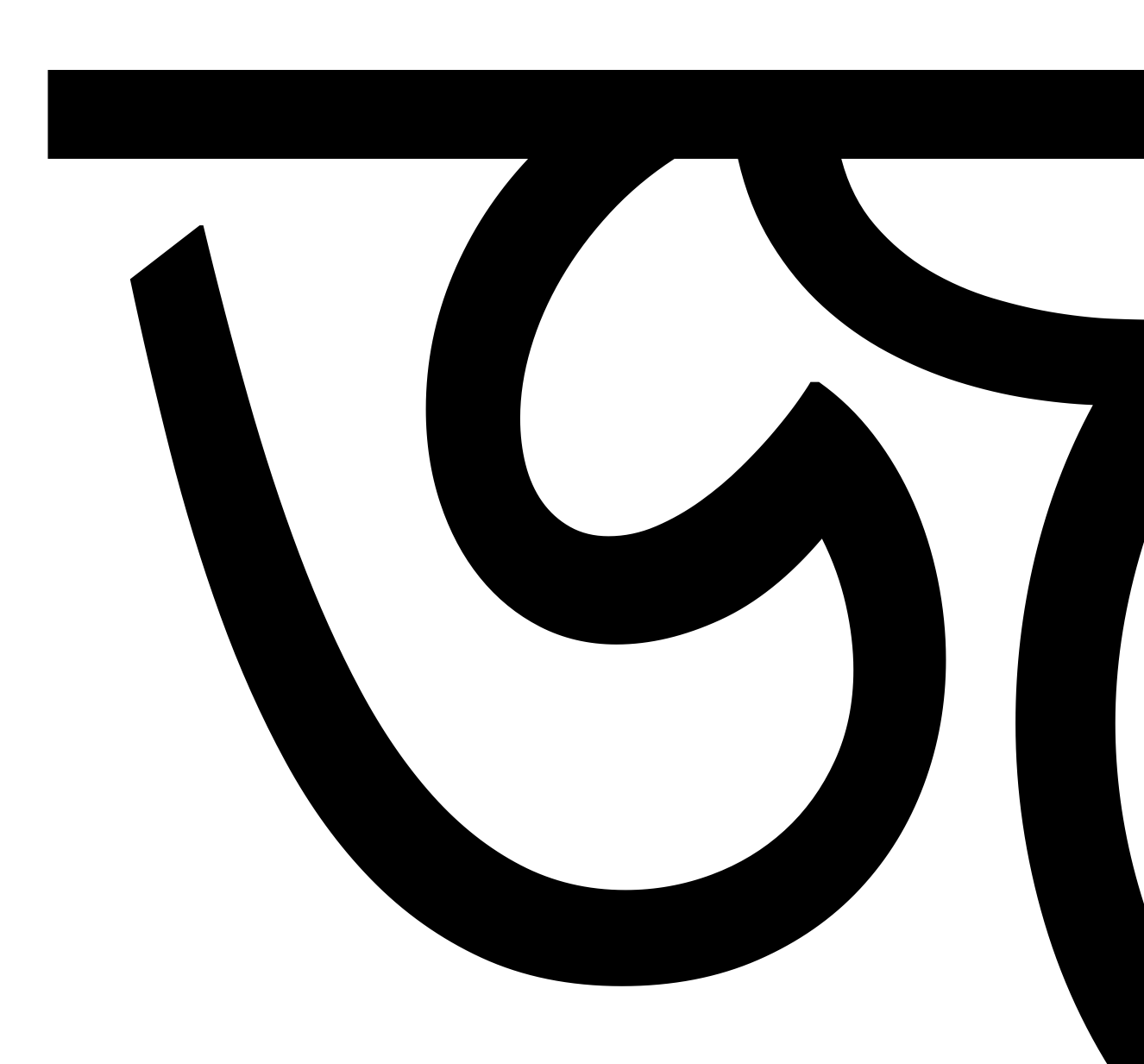}~"
is pronounced as ``j" (as in ``jug"), and is matched with the Bengali phoneme \textipa{/d\textyogh/}.
For English, a language with irregular orthography, we
observe that the grapheme alignments for consonants still mostly match,
although the correspondence is weaker for vowels as can be seen in the confusion matrix in Fig.~\ref{fig:cmat_en_in}

\section{Pronunciation in Grapheme Models}
\label{sec:pronunciation}
The languages studied in this paper come from three different regions of India
and vary characteristically in pronunciation and orthography. Additionally,
Indian English has its own distinctive accent which is influenced by the native
language of the speaker. Naturally, for a grapheme recognizer
that excludes the use of lexicon, we would like to understand how well the
pronunciations are learnt. We carry out this exploration using the alignments
generated by the grapheme models.

\subsection{Accents}
\label{ssec:accents}
We investigate the grapheme-to-phoneme correspondence for two dialects
(British and Indian English) using alignments from models trained on
Indian and British English respectively in Fig.~\ref{fig:cmat_en_gb_in}.
Between the two dialects, we see clear differences in the grapheme-phoneme mapping, for instance,
the grapheme ``\textit{w}" gets mapped to the phoneme ``\textit{w}"
for British English, and to the phoneme ``\textit{v}" for Indian English.
Likewise, the grapheme ``\textit{t}" is also mapped to different phonemes for
the two dialects. As these sounds
are characteristic to Indian English, we can confirm that accented pronunciations
are learnt reliably by the grapheme recognizer simply from the training data.

\subsection{Code-switching}
\label{ssec:codeswitching}
Another important characteristic of Indian speech data is the use of words from
multiple languages in the same utterance (code-switching).
To better understand the pronunciation capability of grapheme models in this setup,
we performed a few variants of the experiments reported above. When trained using
graphemes and data from two languages (English and Bengali),
we observed that the phonemes (e.g. \textipa{/d\textyogh/}) were mapped to the correct
graphemes of both the languages (Fig.~\ref{fig:cmat_bn_bd}), demonstrating the ability of the recognizer to learn a sound that may be
represented by graphemes of more than one language.

We also considered how the model behaves when constrained to the grapheme targets of a single language
while being trained and tested with data from two languages.
In an experiment where training and test data contained a mixture of Latin and Bangla script,
and the model was provided with only the Bangla graphemes,
we observed that the WER (after transliteration normalization) stayed the
same (41.7, row 1 column 2 in Table \ref{table:wer_ce_smbr}), and the model learned to output phonetically
equivalent English words with Bangla script
(e.g. ``\includegraphics[trim=190 120 230 120, height=\fontcharht\font`\B,clip]{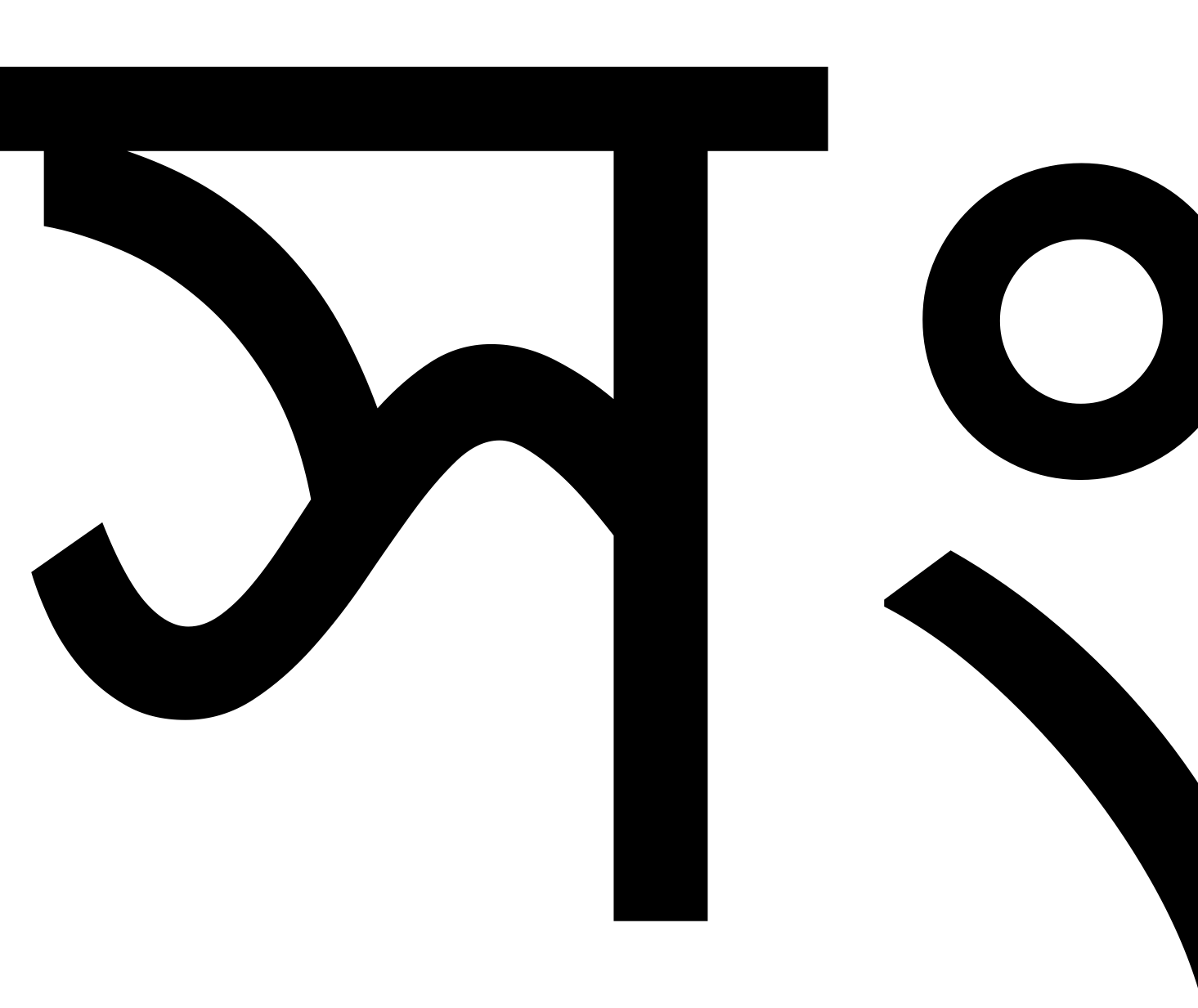}~"
instead of ``song"). These observations led us to believe that the grapheme models were inherently
learning the script-agnostic phonetic pronunciations of the audio, in addition to
the graphemic representations (spelling).

\begin{table}[!t]
\caption{Agreement-score vs G-P WER gap}
\label{table:agreement_score_wer}
\centering
\begin{tabular}{|c||c|c|}
\thickhline
Language     & Agreement-Score & (G-P) WER \\ \hline
Bengali & 46.7\%     & 1.8 \\
Tamil   & 35.9\%     & 1.7 \\
Hindi   & 22.6\%     & 2.5 \\
English & 9.4\%      & 3.4 \\ \thickhline
\end{tabular}
\end{table}

\section{Error Analysis: Grapheme models}
\label{ssec:error_analysis}

Qualitative analysis of the mistakes made by the grapheme models led us to believe that
graphemes that did not have strong correspondence with any specific phoneme were confused
more often. These mistakes surfaced as confusions in similar sounding words (homophones)
or in groups of words. For instance,
the spelling of the ``\textit{u}" sound between ``\textit{stood}" and
``\textit{student}", were frequently interchanged. Similarly, the Bengali grapheme
``\includegraphics[trim=200 120 240 120, height=\fontcharht\font`\B,clip]{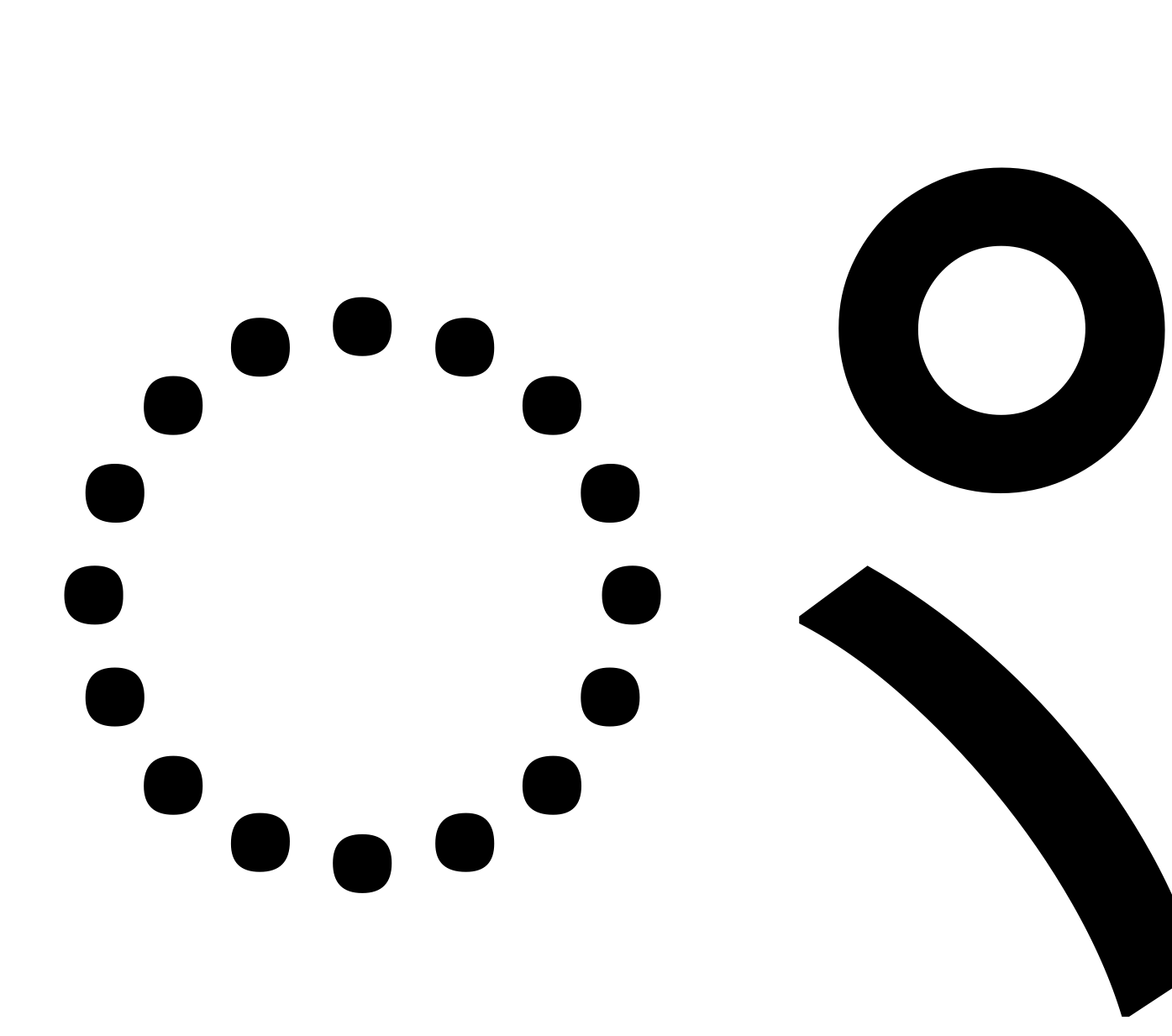}~"
pronounced as ``ng", was replaced with
``\includegraphics[trim=200 120 250 120, height=\fontcharht\font`\B,clip]{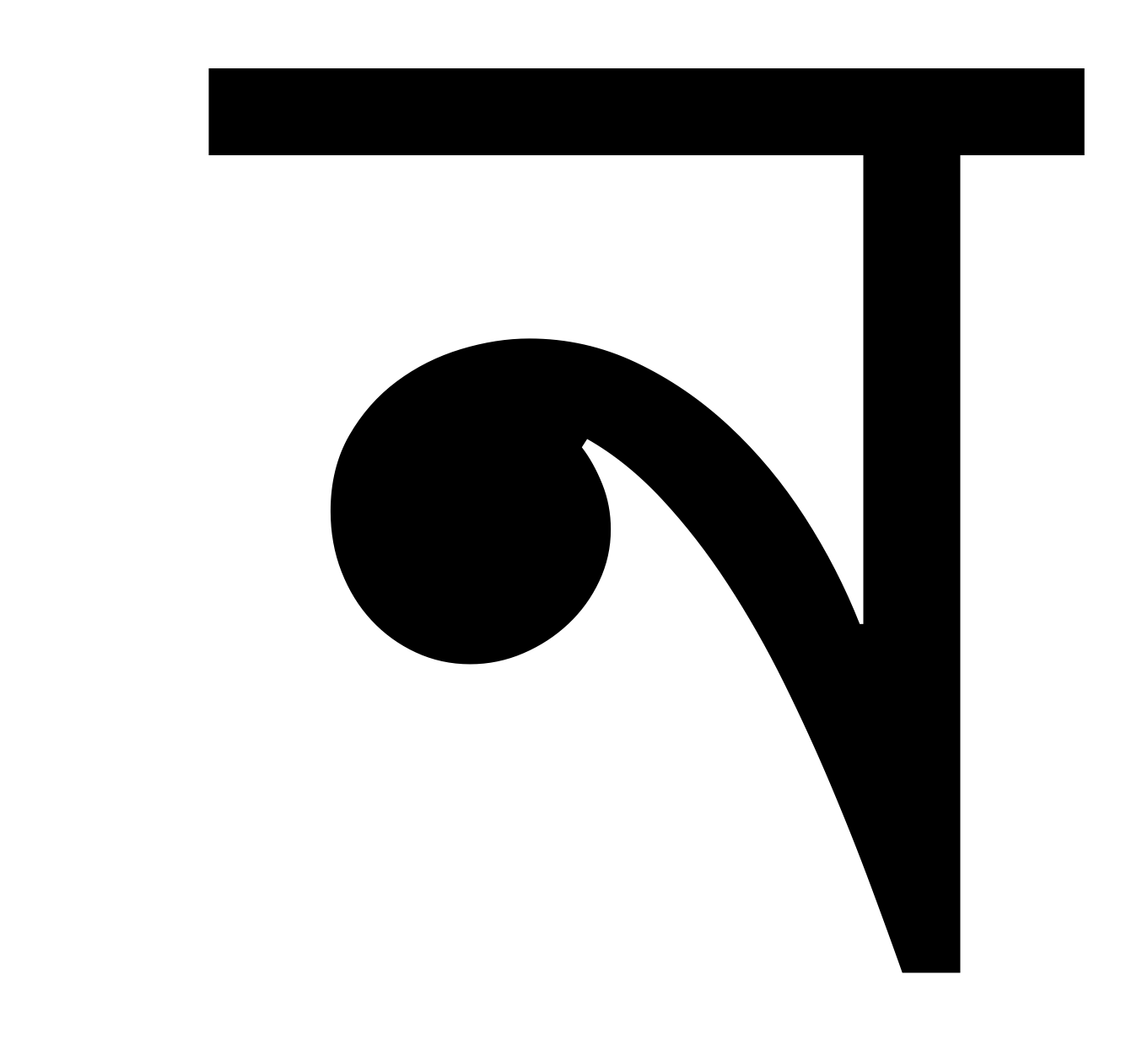}~",
pronounced as ``n" (and vice versa).
To quantify this effect, we computed a metric,
the ``agreement score," which we define as the fraction of graphemes aligning in
at least half of their occurrences with a single phoneme.

\begin{align}
    agreement~score = \frac{\sum_{g\in G}\delta_{g}}{\sum_{g\in G}1}
\end{align}
where
\begin{align}
\delta_{g} =
    \begin{cases}
            1, &         \text{if}~\exists~p\in P, f(g,p) \geq 0.5 ,\\
            0, &         otherwise.
    \end{cases}
\end{align}
for the phoneme set $P$ and grapheme set $G$.
\noindent

Calculating the ``agreement score" on 100K utterances
for all four languages (Table \ref{table:agreement_score_wer}), we observed that the higher
the agreement-score, the lower the WER difference between the Phoneme and Grapheme models.
This suggests that grapheme models might perform better on
languages that have more regular phonemic orthography,
(e.g. Spanish, Finnish, Polish, etc).
Furthermore, we hypothesize that by improving the grapheme-to-phoneme correspondence
in the training data,
we can improve the performance of a grapheme model.

\section{Summary}
\label{sec:summary}

Building speech recognition system for a new language is always a challenge,
especially in acquiring a pronunciation lexicon informed by linguistic knowledge.
In this paper, we propose a simple methodology for training
a grapheme-based recognizer from scratch with efficient training made possible by
hardware acceleration. While graphemic LSTM-based recognizers have been previously explored,
our proposed technique offers the additional advantage of generating good quality audio-to-grapheme alignments
that are valuable for many speech applications such as captioning, keyword spotting and speech synthesis..

Furthermore, using four Indian languages, Bengali, Hindi, Tamil and Indian English with large-scale speech data,
we show that our graphemic models can:
\begin{itemize}
	\item Model the phonetic pronunciations independent of script and orthography.
	\item Capture pronunciations from different accents.
	\item Align a particular phonetic realization with multiple graphemes as seen in code-switched languages.
  \item Achieve performance comparable to lexicon-based models with the same number of parameters.
  \item Be combined with language models and FST decoders for a straightforward
    launch path for applied real-world usage.
\end{itemize}

We believe that our investigation will shed new light on how graphemes can be used
for alternate lexicon-free speech recognition systems.

\section{Acknowledgments}

We would like to thank Tara Sainath, Ehsan Variani, Seungji Lee, Mikaela Grace, and Pedro Moreno for their advice and guidance throughout this project.

\ninept
\bibliographystyle{IEEEtran}
\small\bibliography{refs}

\end{document}